\documentclass{elsart}
\usepackage{graphics}
\usepackage{graphicx}
\usepackage{epsfig}
\usepackage{amssymb}

\begin{document}

\begin{frontmatter}

\title{Competition between ferro-retrieval and anti-ferro orders 
in a Hopfield-like network model for plant intelligence}
\author[label1]{Jun-ichi Inoue}
\ead{j$\underline{\,\,\,}$inoue@complex.eng.hokudai.ac.jp}
\author[label2]{Bikas K. Chakrabarti}
\ead{bikas@cmp.saha.ernet.in}
\address[label1]{Complex Systems Engineering, Graduate School of 
Information Science and Technology, Hokkaido University, 
N13-W8, Kita-Ku, Sapporo 060-8628 Japan}
\address[label2]{Saha Institute of Nuclear Physics, 1/AF Bidhannagar, 
Kolkata-700064, India}

\begin{abstract}
We introduce a simple cellular-network model to explain the capacity of the
plants as memory devices. Following earlier observations 
(Bose \cite{Bose} and others), 
we regard the plant as a network in which each of the elements
(plant cells) are connected via negative (inhibitory) interactions. 
To investigate
the performance of the network, we construct a model following that
of Hopfield, whose energy function possesses both Hebbian spin glass and
anti-ferromagnetic terms. With the assistance of the replica method, we find
that the memory state of the network decreases enormously due to the effect
of the anti-ferromagnetic order induced by the inhibitory connections. We
conclude that the ability of the plant as a memory device is rather weak.
\end{abstract}

\end{frontmatter}

\section{Introduction} \label{intro}
Since the pioneering work by an Indian scientist 
J. C. Bose \cite{Bose}, plants have been regarded as networks which are 
capable of intelligent responses to 
environmental stimuli. 
For example, the dodder coil, which is a plastic plant, 
explores new host tree within hours after their initial touch contact \cite{Trewavas}.
This sort of behavior might be regarded as {\it plant intelligence}. 
If that is the case, 
does the plants compute, learn or memorize various spacial and 
temporal patterns in different environments as computer or our brain does ?  

Recently, Peak et. al. \cite{Peak}
pointed out that 
the plants may regulate their uptake and loss of gases 
by distributed computation. 
As well known, the ability of neural networks, which is a mathematical 
model of brain, is also based on parallel and distributed computation. 
Therefore, similarities 
between neural network model of brains and the plant network should be discussed. 
Although the behavior of the dodder coil we mentioned above is due to 
emergence of the intelligence as a {\it macroscopic function}, 
it is important for us to investigate 
its {\it microscopic reason}. 

Almost eighty years ago, Bose \cite{Bose} detected electrical signaling 
between plants cells. Since his experiments, many examples of 
cross-talk, namely, the biochemical signaling pathways in plants 
have been found. Especially, a Boolean representation of the 
networks of signaling pathways is possible in terms of logical gates 
like AND, OR and XOR etc. These Boolean descriptions make it 
possible to draw analogies between plant networks and neural network models. 

Recently, 
Br$\ddot{\rm u}$ggemann et. al. 
\cite{Bruggemann} found that the plant vacuolar 
membrance current-voltage characteristics is 
quite nonlinear and almost equivalent to 
that of a Zenner diode. Inspired by such observations, 
Chakrabarti and Dutta \cite{Chakrabarti}  identified the cell
membranes as the {\it two state neurons} of the plants and
utilized such a threshold behavior 
of the plant cell membranes to develop or model gates for performing 
simple logical operations. They found that the plant network connections are 
all positive ({\it excitatory}) or all negative 
({\it inhibitory}), compared to the randomly positive-negative 
distributed synaptic connections in real brains. 
As a result, the plant network does not involve any frustrations 
in their computational capabilities. Hence, although the logical gates for
computations could be achieved by such networks of plant cells,
for {\it {intelligence}} of such networks they must must also possess some
memory capacity (in order to caompare and optimise). 

With this fact in mind, in this paper, we 
investigate the equilibrium 
properties of the Hopfield model in which both 
ferromagnetic retrieval and 
anti-ferromagnetic terms co-exsist. 
The strength of the anti-ferromagnetic order is 
controlled by a single parameter $\lambda$. 
Within the replica symmetric calculation, 
we obtain phase diagrams of the system. 
The $\lambda$-dependence of the 
optimal loading rate $\alpha_{c}$ at zero temperature ($T=0$) is discussed. 

This paper is organized as follows. 
In the next section, we introduce several 
experiments and observations 
about the current-voltage characteristics of the plant 
cell membrance. 
In section 3, 
we model the plant with such properties by 
using a Hopfield-like network model 
in which both ferro-retrieval and anti-ferro terms exist. 
In section 4, 
we analyze the model 
with assistance with replica method. 
In this section, we investigate to what extent 
the ferro-magnetic retrieval order 
remain against the anti-ferro disturbance. 
We also investigate the result of the ferromagnetic disturbance. 
In the final section is summary. 
\section{The I-V characteristics of cell membrances}\label{cell}
\begin{figure}[ht]
\begin{center}
\includegraphics[width=7cm]{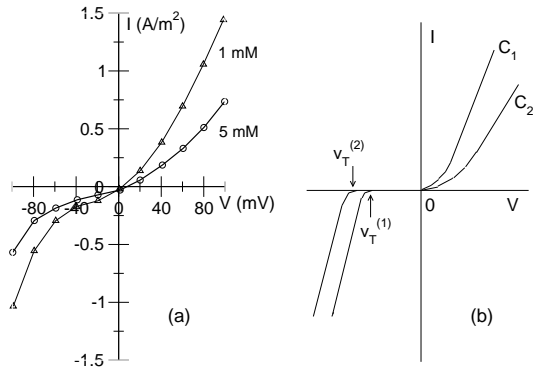}
\end{center}
\caption{\footnotesize 
(a) The non-linear I-V characteristics of cell 
membranes (cf. \cite{Bruggemann}). (b) Its Zener diode like
representation with threshold voltages denoted by $v_T$.
}
\label{fig:fig1}
\end{figure}

In this section, we briefly mention several results concerning 
properties of the plant units, namely, 
current (I) - voltage (V) characteristics of their cell membrane. 
In Fig. \ref{fig:fig1}, we show the typical 
non-liner I-V characteristics of 
cell membranes for the logical gates. 
\vskip 0.3in
From this figure, we find that 
the I-V characteristics is equivalent to 
that of Zenner diode. 
From the view point of input-output logical units 
like perceptrons for neural networks, 
the output of $i$th unit $O_{i}$ is given by \cite{Chakrabarti} 
\begin{eqnarray}
O_{i} & = & \Theta \left(
\sum_{j=}^{N}w_{ij}I_{j} - \theta
\right); ~~ \theta \equiv v_T,
\end{eqnarray}
where the connections $w_{ij}$ are
all positive or all negative. 
In the Hopfield model they are  $\pm$ randomly 
distributed weight matrix as given by  the Hebb rule. 
From these experimental results and simple observations, 
we now have a natural question:   
could the plants act as memory devices as a real brain does ? 
Obviously, 
in the above definition of a single unit of the plant network, 
there is no frustration as in the animal brains. 
Thus, a scope of this paper is to make clear 
this problem, that is to say, 
to what extent, this kind of limitations for sign of the 
weight matrix influences the ability of pattern retrieving as 
associative memories. 

For this purpose, we introduce 
a simplest plant intelligence model based on 
a Hopfield-like model in which 
ferromagnetic retrieval and anti-ferromagnetic 
ordered phases co-exists. 
In the next section, we explain its details.
\section{The plant intelligence model}\label{model}
We start from the following Hamiltonian : 
\begin{eqnarray}
\mathcal{H} & = & 
\frac{1}{N}\sum_{ij}
\left(
\lambda - \sum_{\mu=1}^{p}
\xi_{i}^{\mu}\xi_{j}^{\mu}
\right)S_{i}S_{j} 
\equiv  \mathcal{H}_{\rm AF} + \mathcal{H}_{\rm FR}
\end{eqnarray}
\begin{eqnarray}
\mathcal{H}_{\rm AF} \equiv 
\frac{\lambda}{N}\sum_{ij}S_{i}S_{j},\,\,
\mathcal{H}_{\rm FR} \equiv
-\frac{1}{N}\sum_{ij \mu}
\xi_{i}^{\mu}\xi_{j}^{\mu}
S_{i}S_{j}
\end{eqnarray}
where 
$\mbox{\boldmath $\xi$}^{\mu}=
(\xi_{1}^{\mu},\cdots,\xi_{N}^{\mu})$ is 
$\mu$th embedded pattern and 
$\mbox{\boldmath $S$}=(S_{1},\cdots,S_{N})$ 
means neuronal states. 
A single parameter $\lambda$ 
determines 
the strength of the anti-ferromagnetic order, 
that is to say, 
in the limit of $\lambda \rightarrow \infty$, 
the system is completely determined by $H_{\rm AF}$. 
On the other hand,  
in the limit of $\lambda \rightarrow 0$, 
the system becomes identical to 
the conventional Hopfield model. 
The purpose of this paper is 
to investigate the 
$\lambda$-dependence of the system, 
namely, 
to study the $\lambda$-dependence of 
the optimal loading rate $\alpha_{c}(\lambda)$ at 
$T=0$ by using the technique of 
statistical mechanics for spin glasses. 
\section{Replica symmetric analysis}\label{RS}
In order to evaluate macroscopic properties of the 
system, we first evaluate the averaged free energy  :
\begin{eqnarray}
\ll \log Z \gg_{\mbox{\boldmath $\xi$}} & = & 
\ll \log {\rm tr}_{\{S \}} 
{\rm e}^{-\beta H} \gg_{\mbox{\boldmath $\xi$}} =
\lim_{n \rightarrow 0}
\frac{\ll Z^{n} \gg_{\mbox{\boldmath $\xi$}}-1}
{n}.
\end{eqnarray}
where $\ll \cdots \gg$ 
means the quenched average over the 
$p=N\alpha$ patterns. 
To carry out this average and 
spin trace,  we use the replica method 
\cite{Hertz,Nishimori}. 

After standard algebra \cite{Hertz,Nishimori}, 
we obtain the pattern-averaged replicated partition function as 
follows. 
\begin{eqnarray}
\mbox{} & \mbox{} & 
\ll Z^{n} \gg_{\mbox{\boldmath $\xi$}} = 
\prod_{\alpha, \mu}
\int_{-\infty}^{\infty}
\frac{dM_{\mu}^{\alpha}}
{\sqrt{2\pi/N\beta}}
\int_{-i\infty}^{+i\infty}
\frac{dm_{\alpha}}
{i\sqrt{2\pi/N\beta\lambda}}\\
\mbox{} & \times & 
\int_{-\infty}^{\infty}
dq_{\alpha\beta}
\int_{-i\infty}^{+i\infty}
\frac{dr_{\alpha\beta}}{2\pi i}
{\exp}[-Nf(\mbox{\boldmath $m$}, 
\mbox{\boldmath $q$}, 
\mbox{\boldmath $M$},
\mbox{\boldmath $r$})]. 
\end{eqnarray}
By assuming the replica symmetric ansatz, namely, 
$M_{\mu}^{\alpha}=M,\,\,\,
m_{\alpha}=m,\,\,\,
q_{\alpha\beta}=q,\,\,\,
r_{\alpha\beta}=r$, 
we obtain the free energy density per replica number $n$ as follows. 
\begin{eqnarray}
\mbox{} & \mbox{} & \frac{f(m,q,M,r)}{n} =  
\frac{\beta}{2}M^{2}-
\frac{\beta \lambda}{2}m^{2}+
\frac{\alpha \beta^{2} r}{2}(1-q) \nonumber \\
\mbox{} & + & 
\frac{\alpha}{2}
\left\{
\log [1-\beta(1-q)] - 
\frac{\beta q}{1-\beta(1-q)}
\right\} \nonumber \\
\mbox{} & - & 
\log \int_{-\infty}^{\infty}
Dz\, 
\log 2 \cosh 
\beta(\lambda m + 
\sqrt{\alpha r}z + M) 
\end{eqnarray}
where we defined $Dz \equiv dz {\rm e}^{-z^{2}/2}/\sqrt{2\pi}$. 
We should keep in mind that 
physical meanings of $m$ and $M$ are 
magnetization of the system, overlap between 
the neuronal state $\mbox{\boldmath $S$}$ and 
a specific recalling pattern 
$\mbox{\boldmath $\xi$}^{1}$ among 
$\alpha N$ embedded patterns, respectively. 
$q$ means spin glass order parameters.

In the next section, we evaluate the saddle point of this 
free energy density $f$ and draw phase diagrams to 
specify the pattern retrieval properties of the system. 
\section{Phase diagrams}\label{pd}
In this section, we investigate 
the phase diagram of the system 
by solving the saddle point equations. 

\emph {Saddle point equations:}
By taking the derivatives of $f$ with 
respect to $M,m,r$ and $q$, 
we obtain the saddle point equations.
\begin{eqnarray}
M & = & 
\int_{-\infty}^{\infty}
Dz\, 
\tanh 
\beta [(1-\lambda)M+z\sqrt{\alpha r}] = -m
\label{eq:sp_m} \\
q & = & 
\int_{-\infty}^{\infty}
Dz\, 
\tanh^{2}
\beta [(1-\lambda)M+z\sqrt{\alpha r}]  
\label{eq:sp_q} \\
r & = & 
\frac{q}{[1-\beta(1-q)]^{2}}
\label{eq:sp_r}
\end{eqnarray}
We solve the equations numerically to obtain 
the phase diagram. 

\emph{$T=0$ noise-less limit: }
We first investigate the $T=0$ limit. 
In this limit, obviously, $q \rightarrow 1$. 
After some algebra, we find that  
the optimal loading rate $\alpha_{c}$ 
is determined 
by the point at which the solution of the following 
equation with respect to $y$ vanishes. 
\begin{eqnarray}
y \left\{
\sqrt{\alpha} + 
\sqrt{\frac{2}{\pi}}
(1-\lambda) \,{\rm e}^{-\frac{y^{2}}{2}}
\right\} 
\mbox{}=  
(1-\lambda)
\left\{
1-2H(y)
\right\} 
\end{eqnarray}
where $H(x)$ is defined by 
$H(x) = \int_{x}^{\infty}Dz$.
\begin{figure}[ht]
\begin{center}
\includegraphics[width=7cm]{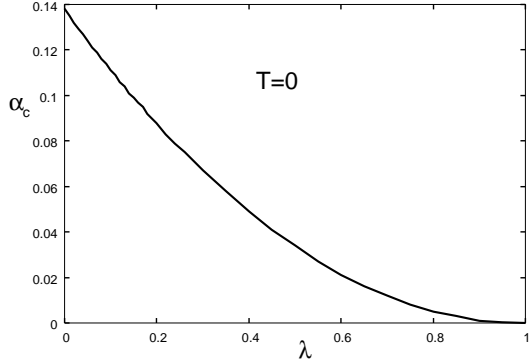}
\end{center}
\caption{\footnotesize 
The optimal loading rate $\alpha_{c}$ as 
a function of $\lambda$. 
$\alpha_{c}(\lambda)$ decreases monotonically.}
\label{fg:fig2}
\end{figure}
In Figure \ref{fg:fig2}, 
we plot the optimal loading rate 
$\alpha_{c}$ as a function of 
$\lambda$. 
From this figure, we see that 
the optimal loading capacity  
$\alpha_{c}(\lambda)$ 
monotonically decreases to zero as $\sim (1-\lambda)^2$. 
This means that 
the ferro-magnetic retrieval order was 
destroyed by adding the anti-ferromagnetic 
term to the Hamiltonian. 
Thus, we conclude that 
if the weight matrix of the 
networks is all positive, 
the plant intelligence model 
does not act as a memory device. 

\emph{Spin Glass-Para phase boundary: }
Before  we solve the saddle point equations for 
$T \neq 0$, it should be important to 
determine the phase boundary between 
the spin glass and para-magnetic phases. 
The phase transition between 
these two phases is 
first order, by expanding 
the saddle point equations 
around $M=q=0$, 
we obtain  
\begin{eqnarray}
q & \simeq & 
\beta^{2}\alpha r \int_{-\infty}^{\infty}z^{2}Dz = \beta^{2}\alpha 
\frac{q}{(1-\beta)^{2}}.
\end{eqnarray}
By solving this equation with scaling 
$\beta \rightarrow (1-\lambda) \beta$ and $T=\beta^{-1}$, 
we obtain the phase boundary line :
\begin{eqnarray}
T_{\rm SG} & = & (1-\lambda)(1+\sqrt{\alpha}) 
\end{eqnarray}

\emph{Phase diagrams for $T\neq 0$: }
Here, 
we investigate the phase diagram for $T\neq 0$ 
by solving the saddle point equations 
(\ref{eq:sp_m}),(\ref{eq:sp_q}) and (\ref{eq:sp_r}) 
numerically. 
\begin{figure}[ht]
\begin{center}
\includegraphics[width=6.5cm]{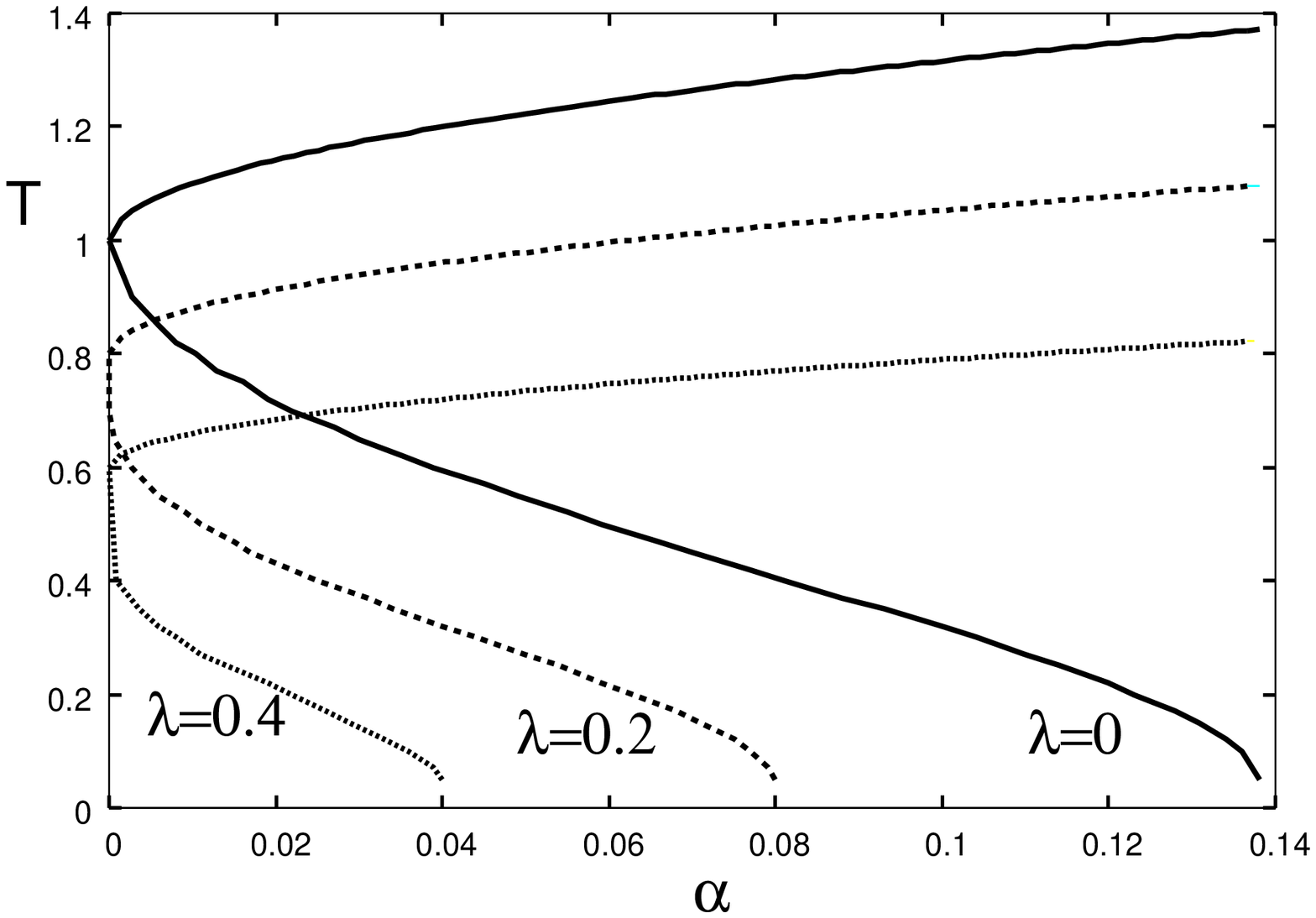}
\includegraphics[width=6.5cm]{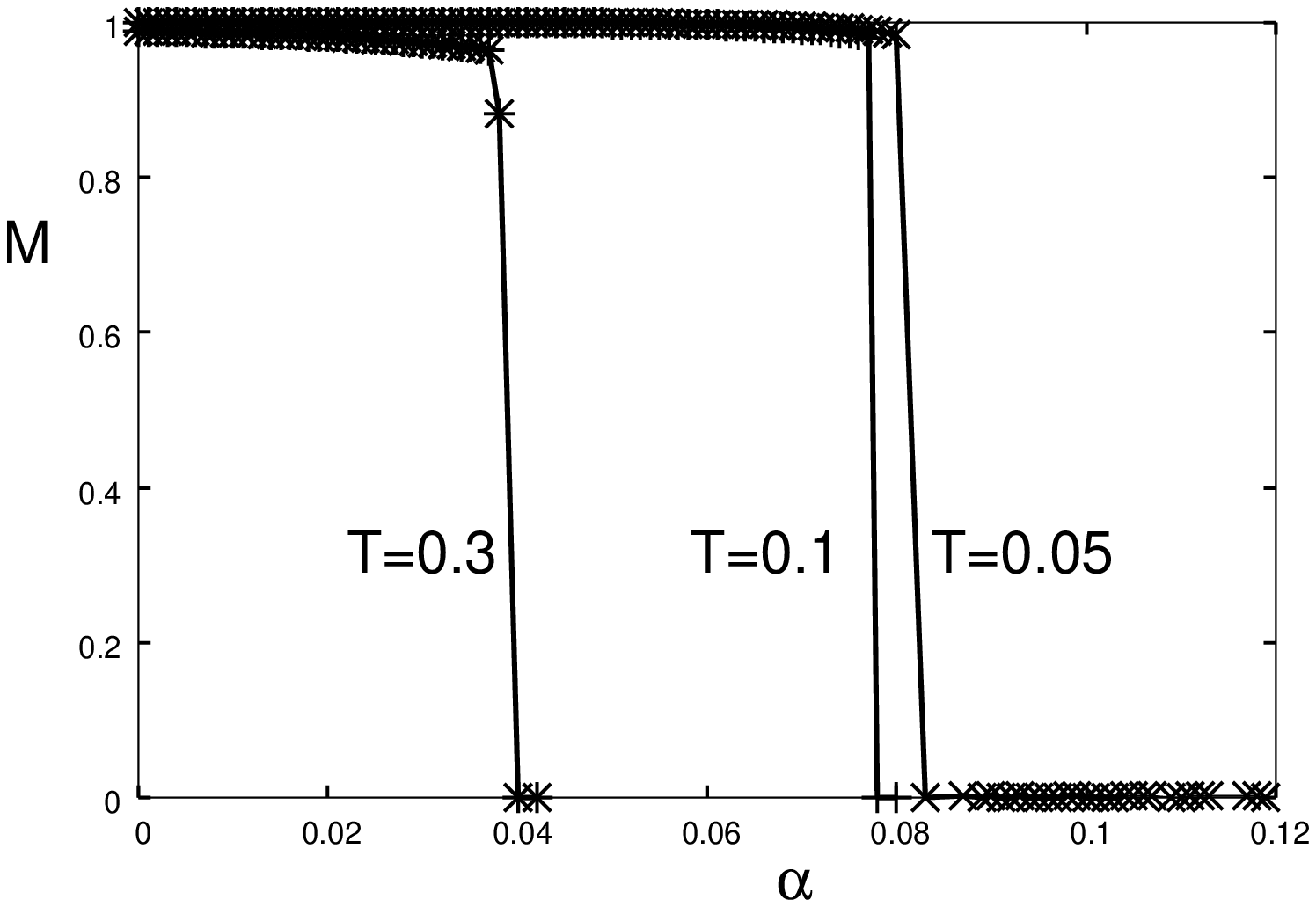}
\end{center}
\caption{\footnotesize  
The phase diagram of the system. 
Ferro-retrieval phase shrinks to 
zero as $\lambda$ increases to $1$.
The para-spinglass boundary is analytically obtained as 
$T_{c}=(1-\lambda)(1+\sqrt{\alpha})$ and is independent of $\lambda$.}
\label{fg:fig3}
\caption{\footnotesize 
The overlap $M$ as a function of $\alpha$ for the case of 
$\lambda=0.2$ at temperatures 
$T=0.05,0.1$ and $T=0.3$.  
}
\label{fg:fig4}
\end{figure}
From this figure, 
we find that 
the ferro magnetic retrieval phase shrinks to 
zero as the anti-ferro magnetic order increases, 
namely, $\lambda \rightarrow 1$. 
The behavior of the overlap $M$ as a function of 
$\alpha$ is shown in the right panel in Figure \ref{fg:fig3}. 
The overlap $M$ becomes zero discontinuously at $\alpha=\alpha_{c}$. 
\begin{figure}[ht]
\begin{center}
\includegraphics[width=6.5cm]{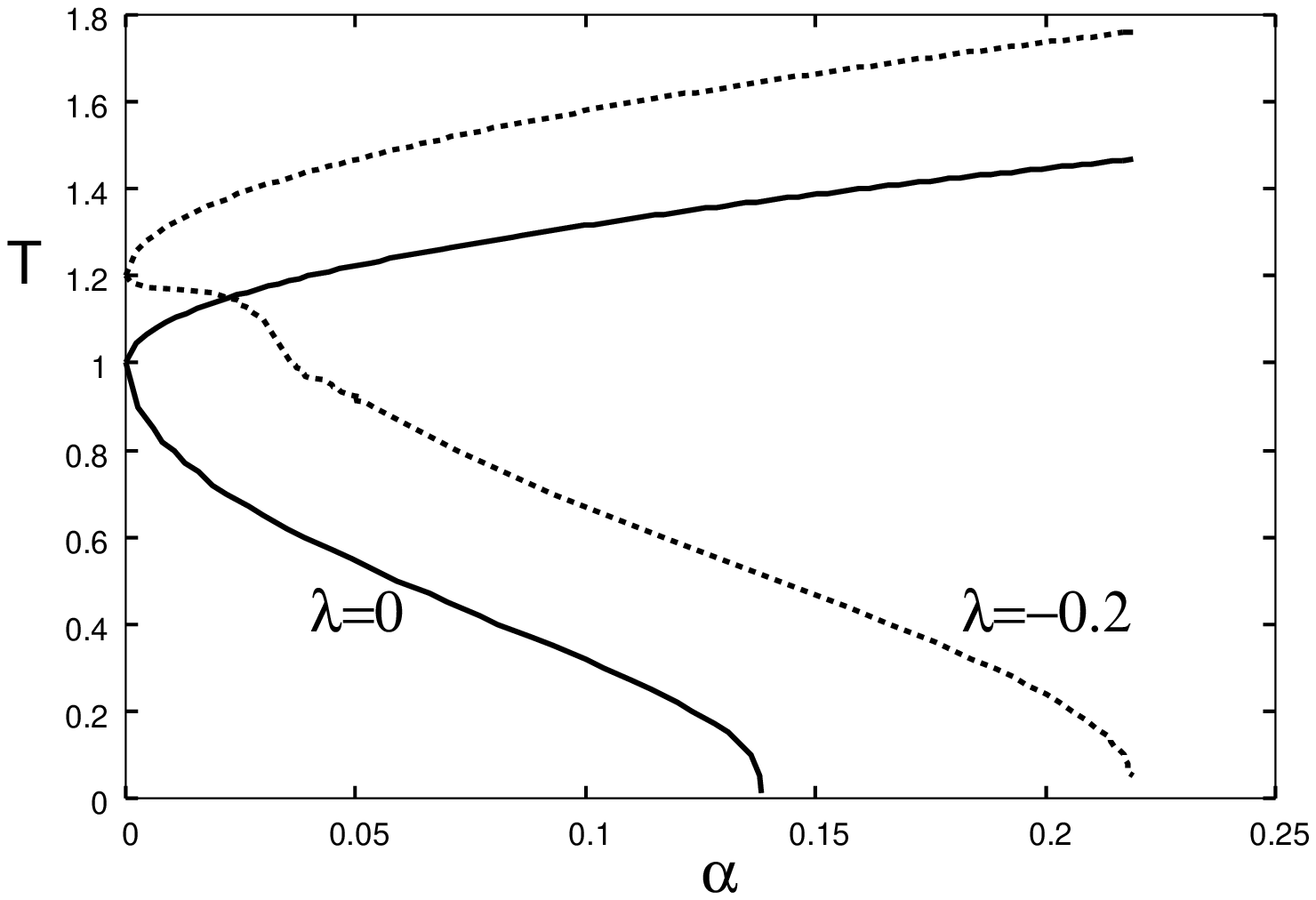}
\includegraphics[width=6.5cm]{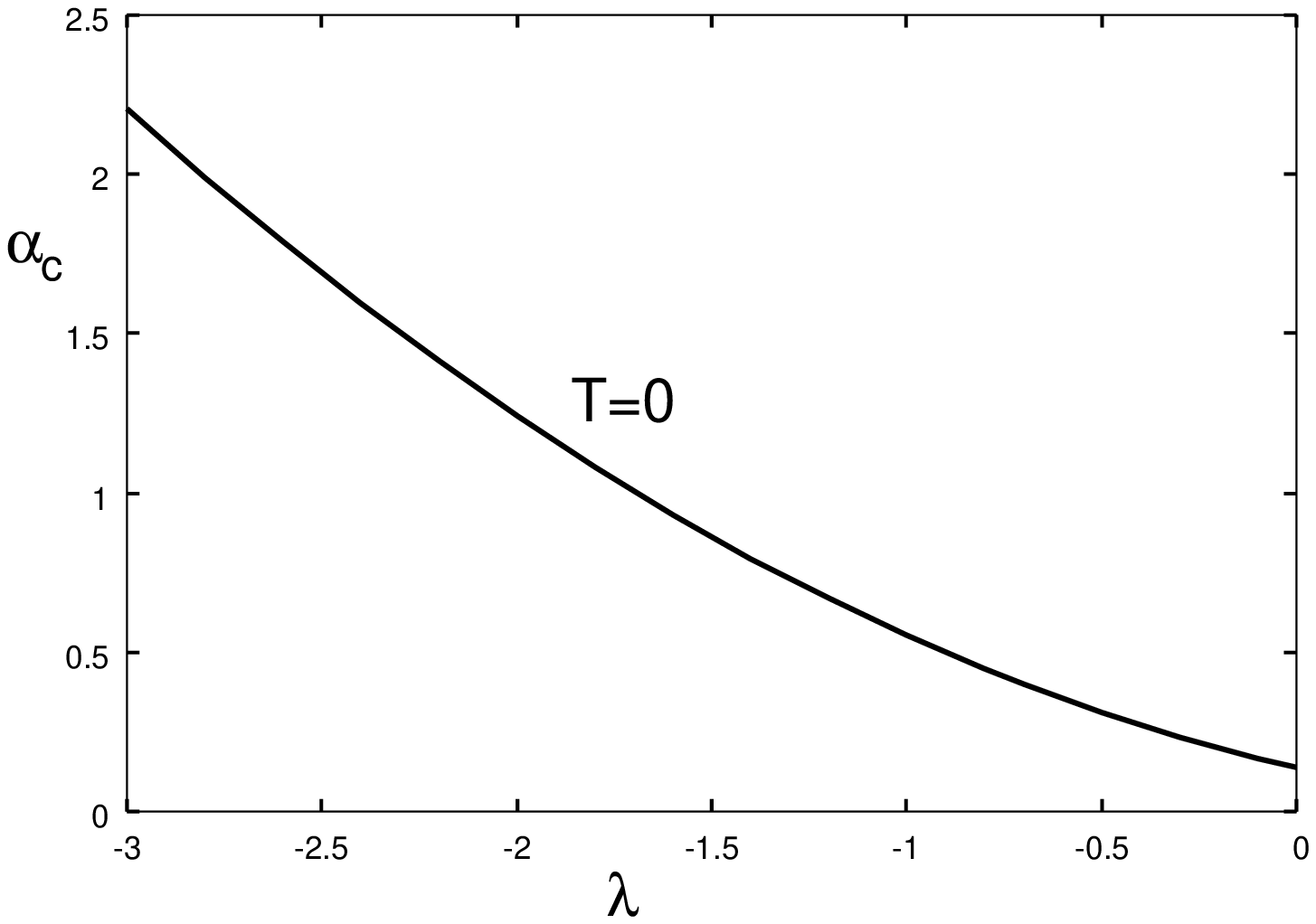}
\end{center}
\caption{\footnotesize  
The phase diagram of the system for negative 
$\lambda=-0.2$.}
\label{fg:fig5}
\caption{\footnotesize  
The optimal loading late $\alpha_{c}$ as a function 
of $\lambda\,(<0)$
at $T=0$.}
\label{fg:fig6}
\end{figure}

\emph{Negative $\lambda$ case: }
Here, we consider the case of negative $\lambda$. 
From the Hamiltonian, we find 
\begin{eqnarray}
\mathcal{H} & = & 
-\frac{1}{N}
\sum_{ij}\sum_{\mu=1}^{p}
\xi_{i}^{\mu}\xi_{j}^{\mu}
S_{i}S_{j} - 
\frac{\lambda^{'}}{N}
\sum_{ij}S_{i}S_{j}; \quad
\mbox{} \lambda^{'}  =  -\lambda \, (> 0). 
\end{eqnarray}
When $\lambda$ increases, the system changes to 
the pure ferromagnet. 
Let us think about 
the limit of $\lambda \rightarrow -\infty$ in the saddle point equation (\ref{eq:sp_m}). 
Then, the term $(1-\lambda)M$ appearing in the argument of 
${\tanh}[\beta (\cdots)]$ becomes dominant, namely, 
$(1-\lambda)M \gg z\sqrt{\alpha r}$ even if 
the loading rate $\alpha$ is large. 
Consequently, equation (\ref{eq:sp_m}) leads to 
\begin{eqnarray}
M & \simeq & 
\int_{-\infty}^{\infty}
Dz \,{\tanh}[\beta (1-\lambda) M] 
\mbox{}  =  \tanh [\beta (1-\lambda) M]. 
\end{eqnarray}
If the factor $1-\lambda$ 
is large enough, the term 
$\tanh [\beta (1-\lambda)M]$ becomes ${\rm sgn}[\beta M]$ and 
saddle point equation (\ref{eq:sp_m}) leads to 
$M  =  {\rm sgn}[\beta M]$.
Apparently, this equation has always a positive solution 
even if the temperature $T=\beta^{-1}$ is large. 
In this sense, 
the factor $(1-\lambda)$ has a meaning of temperature re-scaling. 
It is also possible for us to understand this result 
from the different point of view. In the saddle point equation :
\begin{eqnarray}
M & = & 
\int_{-\infty}^{\infty}
Dz \, \tanh \beta [(1-\lambda)M + z\sqrt{\alpha r}];
\end{eqnarray}
the second term appearing 
in the argument of $\tanh$, 
$z\sqrt{\alpha r}$ means 
{\it cross-talk noise} from 
the other patterns $\mbox{\boldmath $\xi$}^{\mu}$, 
$(\mu=2,\cdots,p)$  and 
obeys Gaussian distribution 
${\rm e}^{-z^{2}/2}/\sqrt{2\pi}$. 
On the other hand, 
the first term $(1-\lambda)M$ 
represents 
{\it signal} of the retrieval pattern 
$\mbox{\boldmath $\xi$}^{1}$. 
Therefore, 
if the second term 
$z\sqrt{\alpha r}$ is 
dominant, the system cannot retrieve the embedded pattern 
$\mbox{\boldmath $\xi$}^{1}$.
Usually, $r$ in the second term grows rapidly as $T$ increases.
And obviously, 
if $\alpha$ increases, the noise term $z\sqrt{\alpha r}$ also 
increases. 
As a result, the signal part $(1-\lambda)M$ becomes relatively small 
and the system moves from the retrieval phase to the spin glass phase. 
However, 
if $\lambda$ is negative large, 
the signal part is dominant and 
the noise part becomes vanishingly small. 
This is an intuitive reason why the optimal loading rate increases 
for negative $\lambda$. 
In Figure \ref{fg:fig6}, 
we plot the optimal loading late $\alpha_{c}$ as a function 
of $\lambda\,(<0)$
at $T=0$. 
As we mentioned already, 
the optimal loading rate $\alpha_{\rm c}$ 
monotonically increases as $\lambda$ goes to 
$-\infty$. 

\section{Summary}\label{sum}

With the identification \cite {Chakrabarti} of plant cell membranes as 
{\it {neurons}}, and demonstrating \cite {Chakrabarti} the possibilty 
of logical gate
operations in these cellular networks, the discussion on
{\it {intelligence}} of a
plant could not be complete without the demonstration of memory in
such, either excitorily or inhibitorily connected, cellular networks. 
In this paper, we introduce a simple model based 
on a Hopfiled-like network  for explaining 
the memory capacity of the plant cell network. 
Following the experimental observations \cite{Bruggemann}, 
we construct a Hopfield model in which 
ferromagnetic retrieval and 
anti-ferromagnetic terms co-exist. 
The strength of disturbance of pattern retrieval by the 
anti-ferro order is controlled by a single parameter: one can easily
see that the internal fields $h_{AF} $ and $h_{FR}$ contributed
by $\mathcal{H}_{AF}$ and $\mathcal{H}_{FR}$ respectively are of order 
$\lambda$ and $\sqrt {pN}/N $ = $\sqrt {\alpha}$ in (3).  
We find that 
the anti-ferromagnetic order 
prevents the system from recalling a pattern. However, even when $\lambda$ 
is greater than $\sqrt {\alpha}$, the network still possess considerable
memory capacity (see Fig. 2).
This result means that the ability of 
the plant as a memory device is rather  weak, if we 
set all weight connections to positive values. 
Our analysis here has been 
for fully connected networks.  
For real plants, the cell membranes should be located on a
finite dimensional lattice \cite{Koyama} or 
on a scale-free network \cite{Stauffer}. 
Investigations for these situations will be 
made in future. 

One of the authors (JI) was supported by 
Grant-in-Aid for Scientific Research on Priority Areas of 
The Ministry of Education, Culture, Sports, Science and Technology (MEXT) 
No. 14084201. BKC is grateful to the Hokkaido University for sponsoring
his visit, during which the work was initiated.


\begin{thebibliography}{99}

\bibitem{Bose}
J. C. Bose, The Nervous Mechanism of Plants, Longmans, London (1923).

\bibitem{Trewavas}
A. Trewavas, Nature {\bf 415} (2002) 841.

\bibitem{Ball}
P. Ball, \emph {Do plants act like computers?},~~~ 
Nature ~~~Science ~~~~ Update ~~~~Weekly ~~~~ Highlights: 26 
January 2004, 
http://info.nature.com/cgi-bin24/DM/y/eNem0CdLrY0C30Hcz0AW.

\bibitem{Peak}
D. A. Peak, J. D. West, S. M. Messinger and K. A. 
Mott, 
Proceedings of Academy of Science USA {\bf 101} (2004) 
 918-922. 


\bibitem{Bruggemann}
L. I. Br$\ddot{\rm u}$ggemann, 
I. I. Pottosin and G. Sch$\ddot{\rm o}$nkecht, 
The Plant Journal {\bf 16} (1998) 101-105. 

\bibitem{Chakrabarti}
B. K. Chakrabarti and O. Dutta, cond-mat/0210538 (2002); Ind. J. Phys. A
{\bf 77} (2003) 549-551.

\bibitem{Bose_I}
I. Bose and R. Karmakar, Physica Scripta  {\bf T106} (2003) 9-12.


\bibitem{Genoud}
T. Genoud and J.-P. Metraux, 
Trends in Plant Science {\bf 4} (1999) 503-507. 

\bibitem{Hertz}
J. Hertz, A. Krough and R. G. Palmer, 
{\it Introduction to the theory of neural computation}, 
Addison-Wesley Publishing (1991).

\bibitem{Nishimori}
H. Nishimori, 
{\it Statistical Physics of Spin Glasses and Information 
Processing}, Oxford Univ. Press (2001). 

\bibitem{Koyama}
S. Koyama, Phys. Rev. E {\bf 65} (2002) 016124. 


\bibitem{Stauffer}
D. Stauffer, A. Aharony, L. de Fontoura Costa and 
J. Adler, 
Euro. Phys. J. B {\bf 32} (2003) 395-399. 




\end{thebibliography}
\end{document}